\documentclass[]{raa}            % referee version: for submission

\usepackage{graphicx,times}             %for PS/EPS graphics inclusion, new
\usepackage{natbib}
\usepackage{amssymb,amsmath}
\bibpunct{(}{)}{;}{a}{}{,}

\usepackage[a4paper=true,dvipdfm=true,pagebackref=true]{hyperref}
\hypersetup{colorlinks = true, linkcolor = green, anchorcolor = red, citecolor = blue, filecolor = red, pagecolor = red, urlcolor = red}

\begin{document}

\title{Photometric investigation and orbital period analyses of the W UMa binaries FP~Lyn, FV~CVn and V354~UMa.}

   \volnopage{Vol.0 (20xx) No.0, 000--000}      %%preserved for Editor. DOn't remove!
   \setcounter{page}{1}          %%starting page, preserved for Editor. DOn't remove!

\author{ Michel, R.\inst{1}, Xia, Q.-Q. \inst{2} and Higuera, J.\inst{1}}

\institute{Observatorio Astron\'{o}mico Nacional, Instituto de Astronom\'{i}a, Universidad Nacional Aut\'{o}noma de M\'{e}xico, Apartado Postal 877, Ensenada, B.C. 22830, M\'{e}xico\\
  \and
Shandong Provincial Key Laboratory of Optical Astronomy and Solar-Terrestrial Environment, Institute of Space Sciences, Shandong University, Weihai, 264209, China; {\it xiaqi77@126.com}\\
\vs\no
   {\small Received~~20xx month day; accepted~~20xx~~month day}}

\abstract{New light curves and photometric solutions of FP Lyn, FV CVn and V354 UMa are presented. It is found that these three systems are W-subtype shallow contact binaries. In addition, it is obvious that the light curves of FP Lyn and V354 UMa are asymmetric. Therefore, a hot spot was added on the primary star of FP Lyn and a dark spot was added on the secondary star of V354 UMa. At the same time, we added a third light to the photometric solution of FP Lyn for the final result. The obtained mass ratios and fill-out factors are $q=1.153$ and $f=13.4\%$ for FP Lyn, $q=1.075$ and $f=4.6\%$ for FV CVn, $q=3.623$ and $f=10.7\%$ for V354 UMa. The investigations of orbital period of these three systems indicate that the periods are variable. FP Lyn and V354 UMa were discovered existing the secular increase component with a rate of $dp/dt=4.19\times10^{-7}$ d yr$^{-1}$ and $dp/dt=7.70\times10^{-7}$ d yr$^{-1}$ respectively, which are feasibly caused by the conservative mass transfer from the less massive component to the more massive component. In addition, some variable  components were discovered for FV CVn, which is a rate of $dp/dt=-1.13\times10^{-6}$ d yr$^{-1}$ accompanied by a cyclic oscillation with amplitude and period of 0.0069 days and 10.65 yr. The most likely explanation for the long-term decrease is angular momentum loss. And the existence of an additional component is the most plausible explanation for the periodic variation.}

\keywords{stars: binaries: close ---
         stars: binaries: eclipsing ---
         stars: individual (FP Lyn, FV CVn and V354 UMa) }

 \authorrunning{Michel R., Xia Q.-Q. \& Higuera J.}            %author_head in even pages
 \titlerunning{Photometry of FP~Lyn, FV~CVn and V354~UMa}  % title_head in odd pages

 \maketitle

\section{Introduction}
W UMa type stars are eclipsing binaries with orbital periods of usually less than one day. W UMa stars are close binaries of spectral types F, G or K. Both stars are filling or over-filling their Roche lobes, thus sharing a common convective envelope (CCE) (Lucy 1968a, 1968b). W UMa type stars transfer energy between the two components through their CCE, and their strong interaction affect their evolution properties such as mass transfer (e.g., Li et al. 2015a, 2018; Liao et al. 2017), magnetic activity (e.g., Applegate 1992; Yuan \& Qian 2007; Li et al. 2014), thermal relaxation oscillation (TRO) (e.g., Lucy 1976; Flannery 1976; Robertson \& Eggleton 1977) and angular momentum loss (AML) (e.g., Van't Veer 1979; Rahunan 1981; van Hamme 1982; Qian 2001a, 2001b, 2003), which is also very important for studying other eclipsing binaries. The results of orbital period analyses have shown that the presence of a third body is a common feature of W UMa systems. For example, V781 Tau (Li et al. 2016a), V502 Oph (Zhou et al. 2016a), TYC 1337-1137-1 and TYC 3836-0854-1 (Liao et al. 2017), V0474 Cam (Guo et al. 2018). The observations and statistical analyses of W UMa type stars are of great significance for understanding the evolution process of eclipsing binaries (eg., Qian et al. 2017, 2018).

FP~Lyn was detected as an EW type binary star by Maciejewski \& Niedzielski (2005) who reported an orbital period of 0.359097 days and a maximumn V magnitude of 11.36. Later, Hoffman et al. (2009) identified 4659 variable objects in the Northern Sky Variability Survey and classified FP Lyn as a W UMa candidate with an orbital period of 0.35913 days. Recently, H\"{u}bscher (2016) and Jury\v{s}ek et al. (2017) have reported the timing of minima of many eclipsing binaries including those of FP~Lyn. Nevertheless, the study of photometric solutions and the period variations have not been published yet. Therefore, in this paper, the photometric solutions and O-C analyses were obtained by using the new light curves of $BVR_c$ and all minimal timings that have been observed.

FV CVn was firstly identified as an eclipsing binary system whose period is 0.31539899 days from ROTSE all-sky surveys for variable stars published by Akerlof et al.(2000). Then, as a byproduct of the ROTSE1 CCD survey, FV CVn was observed with CCD equipment by Bl\"{a}ttler \& Diethelm (2004). The period was revised as 0.315367 days and proved the object to be a W UMa type binary system. Timings of minima of FV CVn were obtained by several authors later, such as Diethelm (2005, 2006, 2007) and Nelson (2016). However, there has been no analysis about the physical parameters and evolutionary state of FV CVn. Recently, we observed FV CVn for two nights, and the complete light curves of $BVR_cI_c$ were obtained. So, we analyzed the orbital period and the light curves with all available data.

V354 UMa is a W UMa eclipsing binary according to ROTSE-I data (Wozniak et al., 2004). The light curves were firstly obtained by Khruslov (2006), and it further proves that it is a W UMa type binary whose period is 0.293825 days. From Norton et al. (2007), 11.02 mag of V band was determined. Then, some minimal timings were obtained from Nelson (2015) and H\"{u}bscher (2016), etc. And it also has no photometric analyses and period investigations. Therefore, we analyzed its light curves and orbital periods using the new observed $UBVR_cI_c$ light curves and all available minima in this paper.

\section{Observations}
These three objects were observed by the 84cm telescope at the National Astronomical Observatory of Mexico (OAN-SPM) to obtain the CCD observations. Each system was observed for two nights. Standard Johnson-Cousin-Bessel $UBVR_cI_c$ filters were used to observe these three systems. Information about the observation time, filters and exposure time of FP Lyn, FV CVn and V354 UMa are listed in Table 1. One in particular is the different exposure time of FP Lyn in January 31, 2016, we changed the exposure time to obtain the better observations. In order to find the best reference stars in the fields, those with similar color to the variables, the fields were calibrated during very photometric conditions and using Landolt standard stars. For FP~Lyn, star  2MASSJ08420426+3905326 ($U=12.997, B=13.097, V=12.544, R=12.189, I=11.850$) was chosen as reference star while 2MASSJ13533984+3216259 ($U=14.580, B=14.684, V=14.233, R=13.930, I=13.638$) was used in the case of FV~CVn and TYC3466-294-1 ($U=12.373, B=12.286, V=11.723, R=11.413, I=11.105$) for V354~UMa. The light curves of these three objects are displayed in Figure 1. It is obvious that they are all W~UMa type binary systems, and the best way to explain the asymmetric light curves of FP~Lyn and V354~UMa is the existence of O'Connell effect.

\begin{table}
\begin{center}
\caption{Information of Observation for FP Lyn, FV CVn and V354 UMa}
\begin{tabular}{lccc}
\hline\hline
Objects	& Date	& Filter (Exposure Time in seconds) & Type  \\
\hline

%\multirow{2}{*}{FP Lyn}	& \multirow{2}{*}{January 31, 2016}  & $B(45), V(25), R_c(15)$ \& $B(40), V(20),$& \multirow{2}{*}{Complete light curves}\\
FP Lyn	& January 31, 2016  & $B(45), V(25), R_c(15)$ \& $B(40), V(20),$& Complete light curves\\
        &                   & $R_c(10)$ \& $B(20), V(10), R_c(5)$  &        \\
        & February 20, 2016 & $B(32), V(16), R_c(8)$              & Minimum light        \\
FV CVn  & March 30, 2017    & $B(50), V(25), R_c(15), I_c(15)$     & Minimum light        \\
        & April 21, 2018    & $B(30), V(20), R_c(15), I_c(15)$     & Complete light curves\\
V354 UMa& March 31, 2017    & $B(30), V(12), R_c(8), I_c(8)$       & Minimum light        \\
        & April 23, 2018    & $U(40), B(20), V(10), R_c(6), I_c(6)$ & Complete light curves\\
\hline
\end{tabular}
\end{center}
\end{table}

\begin{figure}\centering
\includegraphics[width=0.32\textwidth]{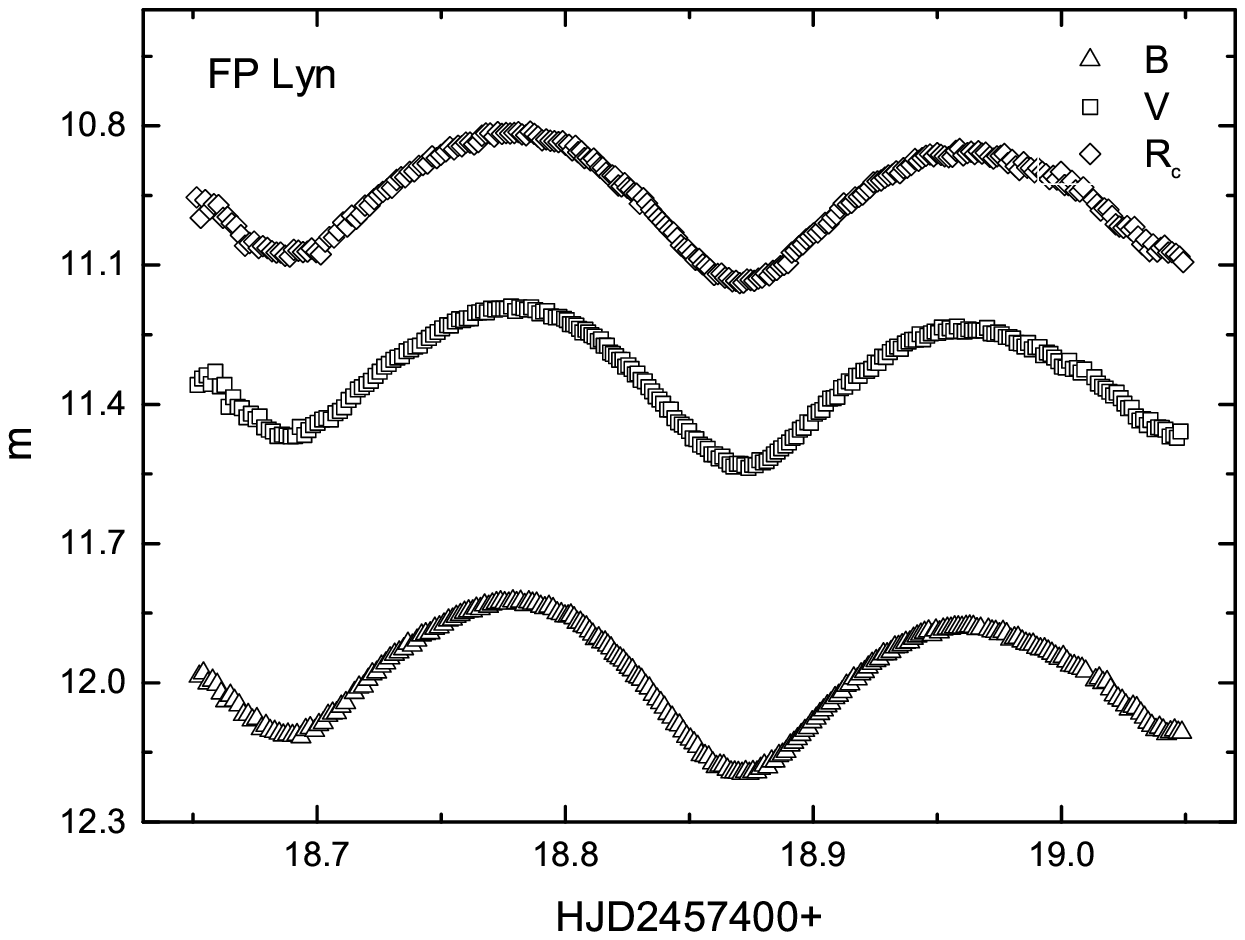}
\includegraphics[width=0.32\textwidth]{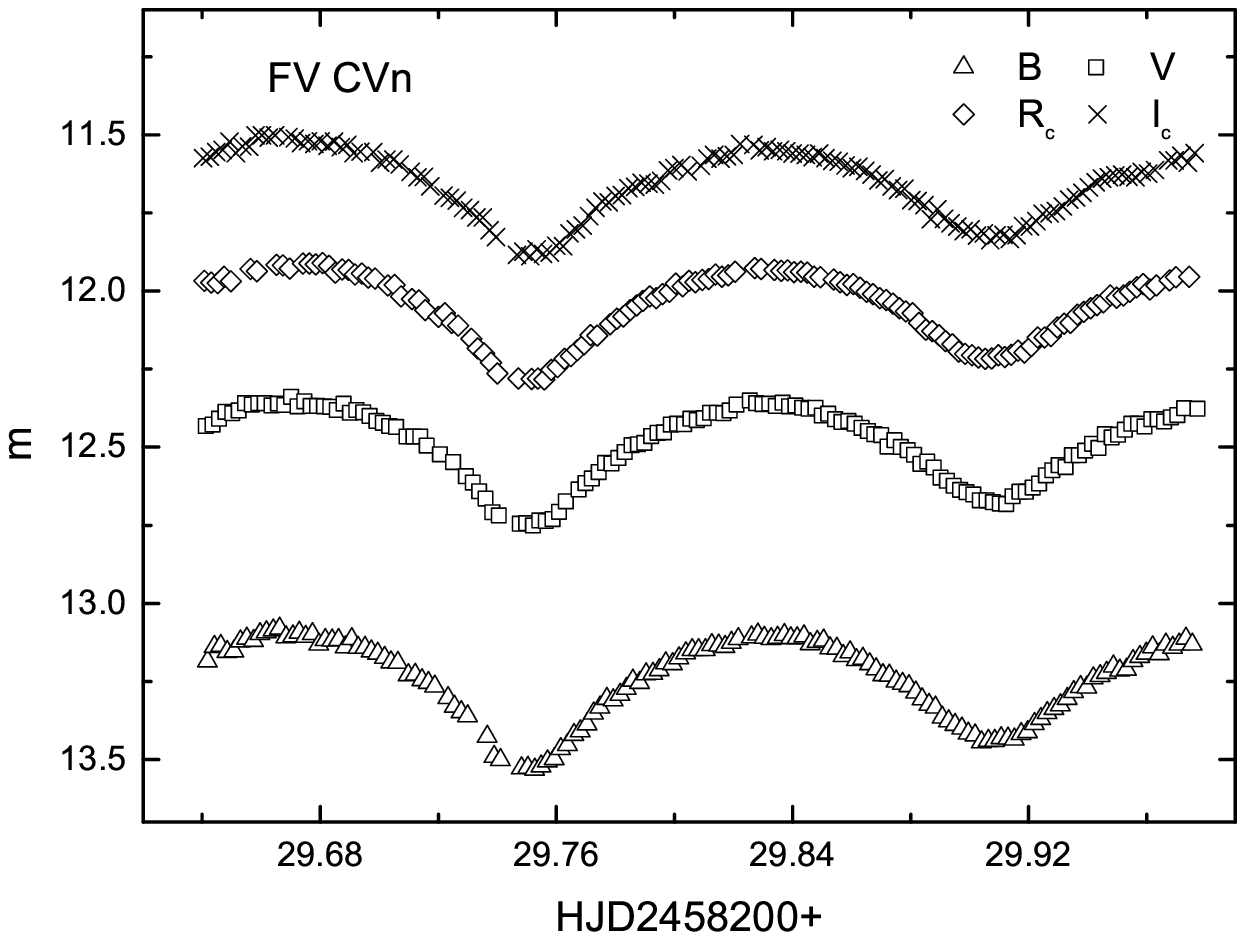}
\includegraphics[width=0.32\textwidth]{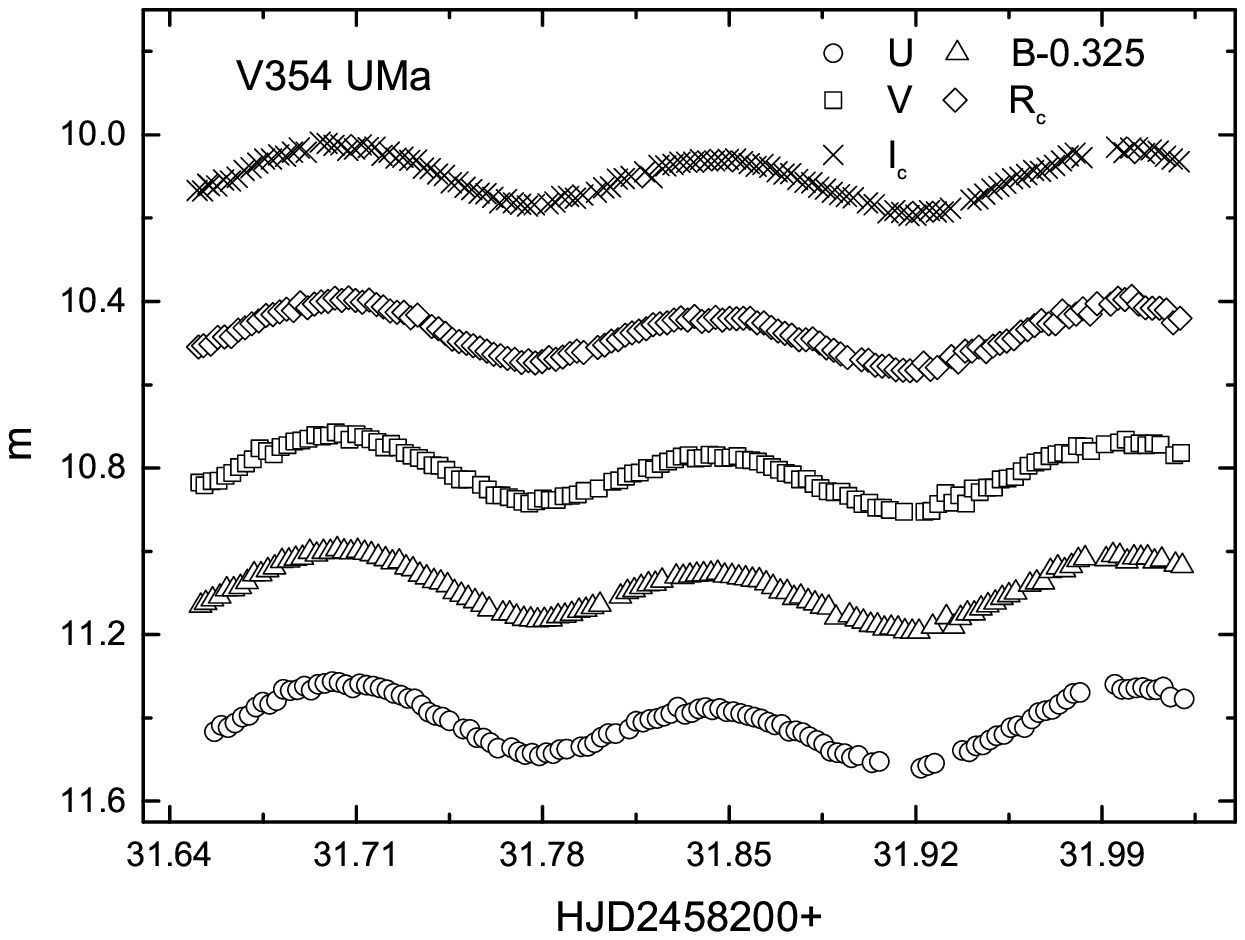}
\caption{CCD photometric light curves of FP Lyn (left panel), FV CVn(middle panel) and V354 UMa (right panel). FP Lyn: $BVR_c$ bands observed on January 31, 2016; FV CVn: $BVR_cI_c$ bands observed on April 21, 2018; V354 UMa: $UBVR_cI_c$ bands observed on April 23, 2018.}
\end{figure}

\section{Photometric solutions}
Wilson-Devinney code of 2013 version was used to analyze the new light curves of FP Lyn, FV CVn and V354 UMa in order to obtain their model parameters. The initial epoch and period used to convert time into phase are as follows,
\begin{equation}
\begin{split}
\label{equation:emphirical}
\textrm{FP Lyn: Min.}I=2457418.6891+0.359085\times E,\\
\textrm{FV CVn: Min.}I=2458229.7505+0.315365\times E,\\
\textrm{V354 UMa: Min.}I=2458231.7783+0.293825\times E,
\end{split}
\end{equation}
where the initial epochs are calculated from our observations, the periods of FP Lyn and FV CVn are from O-C Gateway\footnote{http://var2.astro.cz/ocgate/} and period of V354 UMa is obtained from VSX\footnote{https://www.aavso.org/vsx/}.

During the W-D solutions, the effective temperatures of primary star are from the Guoshoujing Telescope (the Large Sky Area Multi-Object Fiber Spectroscopic Telescope (LAMOST) ) Data Release 5\footnote{http://dr5.lamost.org/} (e.g., Luo et al. 2015). Partial spectral information of these three systems including the effective temperatures are listed in Table 2. Therefore, we averaged their multiple observations respectively. At last, 5907K, 5470K and 5841K were determined as $T_1$ of these three targets. According to Lucy (1967) and Ruci\'{n}ski (1969), the gravity-darkening coefficients and bolometric albedo coefficients were set as $g_{1,2}=0.32$ and $A_{1,2}=0.5$, respectively. Bolometric limb-darkening and bandpass limb-darkening coefficients were obtained from van Hamme (1993). Considering their light curves, it is obvious that they are all W UMa type binaries. Therefore, mode 3 was chosen for all systems during the modeling. The adjustable parameters are: orbital inclination $i$, effective temperature of the secondary star $T_2$, dimensionless potential of two components $\Omega_1=\Omega_2$ and luminosity of primary star $L_1$.

As the photometric investigations of the three targets are absent since their discovery, $q$-search method was used to determine their mass ratios. In order to find the minimum convergent solutions, we set $q$ as fixed values from 0.1 to 10.0, and changed it every 0.1 step. The final residuals and curves corresponding $q$ values are plotted in the left panel of Figures 2, 3, and 4. Then, from the three small insert figures, the $q$ values corresponding to the minimum convergent solutions were determined as 1.4, 1.1 and 3.4, respectively. Therefore, we set them as the initial values of $q$ and adjustable. In the meantime, other adjustable parameters are the same as before. Then the $q$ of convergent solutions were determined as 1.321, 1.071 and 3.299, respectively. In order to obtain better fitting results, dark spot or hot spot model and parameters of the third light were also tried during the W-D solutions.

The light curves of FP Lyn and V354 UMa are both asymmetric which indicates that in both of them exists the O'Connell effect. It is generally explained by the magnetic activity of a dark or hot spot (e.g., AD Cnc, Qian et al. 2007; GSC 03526-01995, Liao et al. 2012; V441 Lac, Li et al. 2016b; UCAC4
436-062932, Zhou et al. 2016b). Therefore, we added a hot spot or dark spot to the primary or secondary star of FP Lyn and V354 UMa, respectively. All fitting parameters are listed in Table 3. Ultimately, the fitting result with the smallest residual is selected as a hot spot on the primary star of FP Lyn and a dark spot on the secondary star of V354 UMa.

Meanwhile, we added the parameters of third light in W-D solutions. FP Lyn and FV CVn were ultimately found the convergent solutions, while no convergent solution was found for V354 UMa. The parameters of the third light were also tabulated in Table 3. It is obvious that the effect of the third body on their total luminosity of FP Lyn cannot be ignored, while FV CVn has too little third light. Therefore, we just chose FP Lyn adding the third light for its final result.

All parameters and errors were determined and listed in Table 3. The errors were obtained by the W-D method. They are all mathematical fitting and have no meaning for physics. The fitting figures of original data and the theoretical light curves were plotted in the right panel of Figures 2, 3 and 4. All residuals (observed minus calculated) of these three systems are almost flat, indicating the theoretical light curves and original data were fitted well. The geometric structure at phase 0.75 of these three systems are displayed in Figure 5.

\begin{table}
\begin{center}
\caption{The spectral information of the three targets released by LAMOST}
\begin{tabular}{lccccc}
\hline
Targets & Obsdate & Subclass & [Fe/H] & $T_{eff}(K)$ & log(g) \\
\hline
FP Lyn & 2014/11/20 & G3 & 0.258 & 5928.24 & 4.182 \\
       & 2014/12/08 & G3 & 0.216 & 5910.85 & 4.202 \\
       & 2014/12/26 & G3 & 0.187 & 5889.37 & 4.138 \\
FV CVn & 2013/04/20 & G7 & 0.006 & 5491.78 & 4.342 \\
       & 2014/01/17 & G7 & 0.064 & 5610.87 & 4.438 \\
       & 2014/03/25 & G7 & 0.057 & 5607.27 & 4.455 \\
       & 2015/04/07 & G7 & -0.154 & 5253.29 & 4.240 \\
       & 2017/02/16 & G5 & -0.099 & 5385.65 & 4.309 \\
V354 UMa & 2015/01/09 & G0 & -0.557 & 5811.61 & 4.153 \\
         & 2017/03/14 & G1 & -0.341 & 5869.98 & 4.324 \\
\hline
FP Lyn &     &    &       & 5909 & \\
FV CVn &     &    &       & 5470 & \\
V354 UMa &    &    &      & 5841 & \\
\hline
\end{tabular}
\end{center}
\scriptsize
The temperature of primary star of FP Lyn, FV CVn and V354 UMa what we used in calculation are 5909K, 5470K and 5841K, respectively, which are the average of their observations.
\end{table}

\begin{figure}
\begin{minipage}[c]{0.5\textwidth}
\begin{center}
\includegraphics[angle=0,scale=0.65]{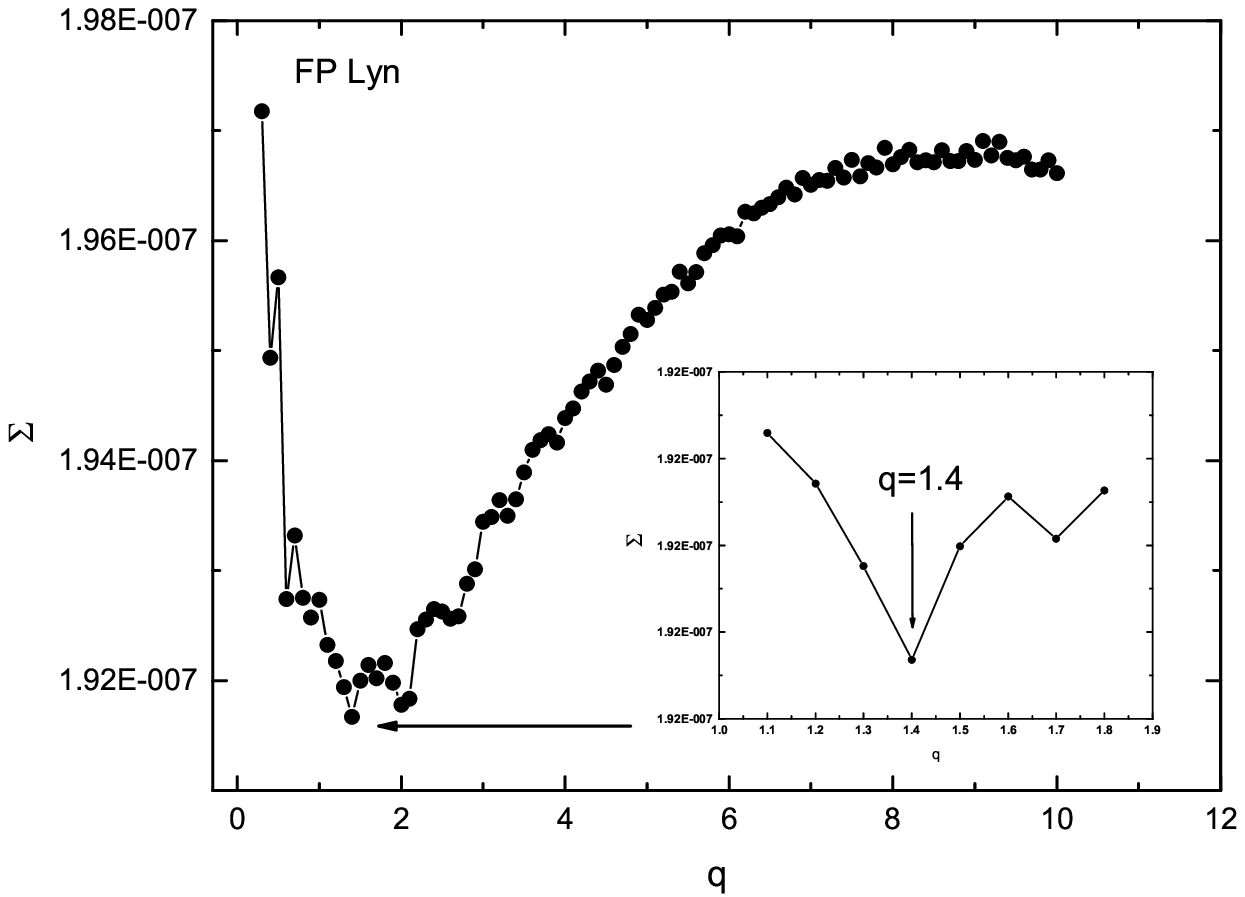}
\end{center}
\end{minipage}
\begin{minipage}[c]{0.5\textwidth}
\begin{center}
\includegraphics[angle=0,scale=0.64]{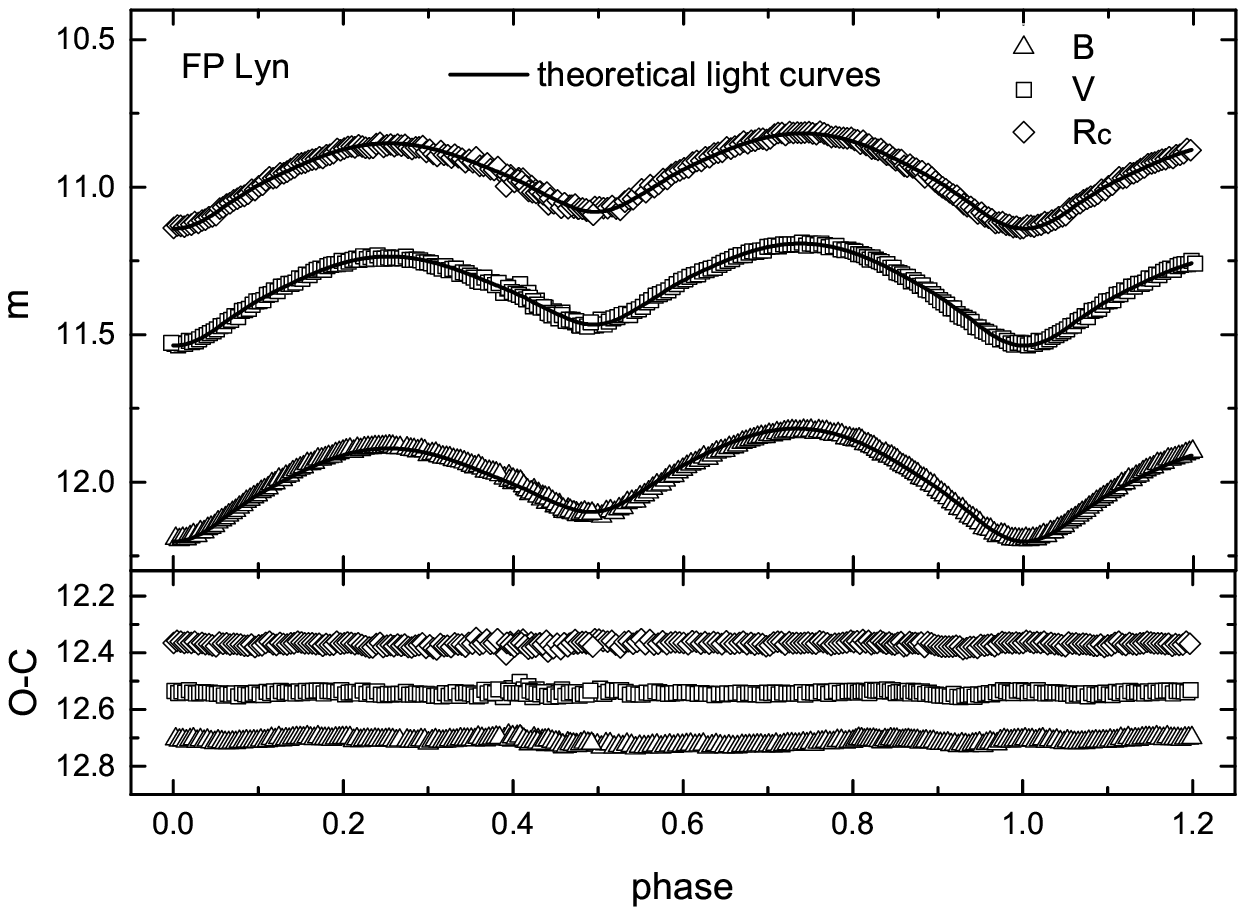}
\end{center}
\end{minipage}
\caption{The left panel is the $\Sigma$ - q curves of FP Lyn. The small insert figure is an enlargement that q is from 1.1 to 1.8. The right panel is the agreement of FP Lyn between observations and theoretical light curves derived by V, $R_c$, and $I_c$ light curves. It was computed with a hot spot on the primary component.}
\end{figure}

\begin{figure}
\begin{minipage}[c]{0.5\textwidth}
\begin{center}
\includegraphics[angle=0,scale=0.65]{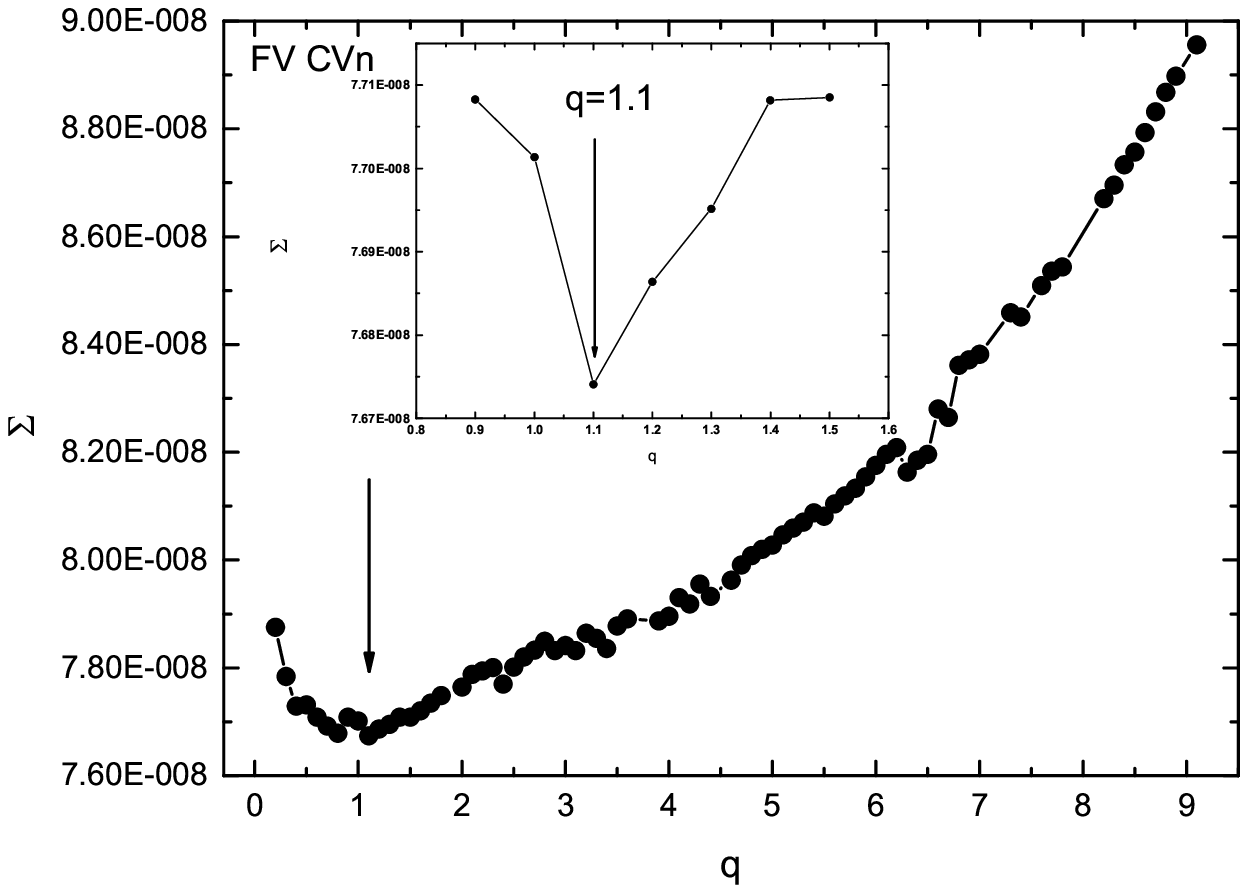}
\end{center}
\end{minipage}
\begin{minipage}[c]{0.5\textwidth}
\begin{center}
\includegraphics[angle=0,scale=0.63]{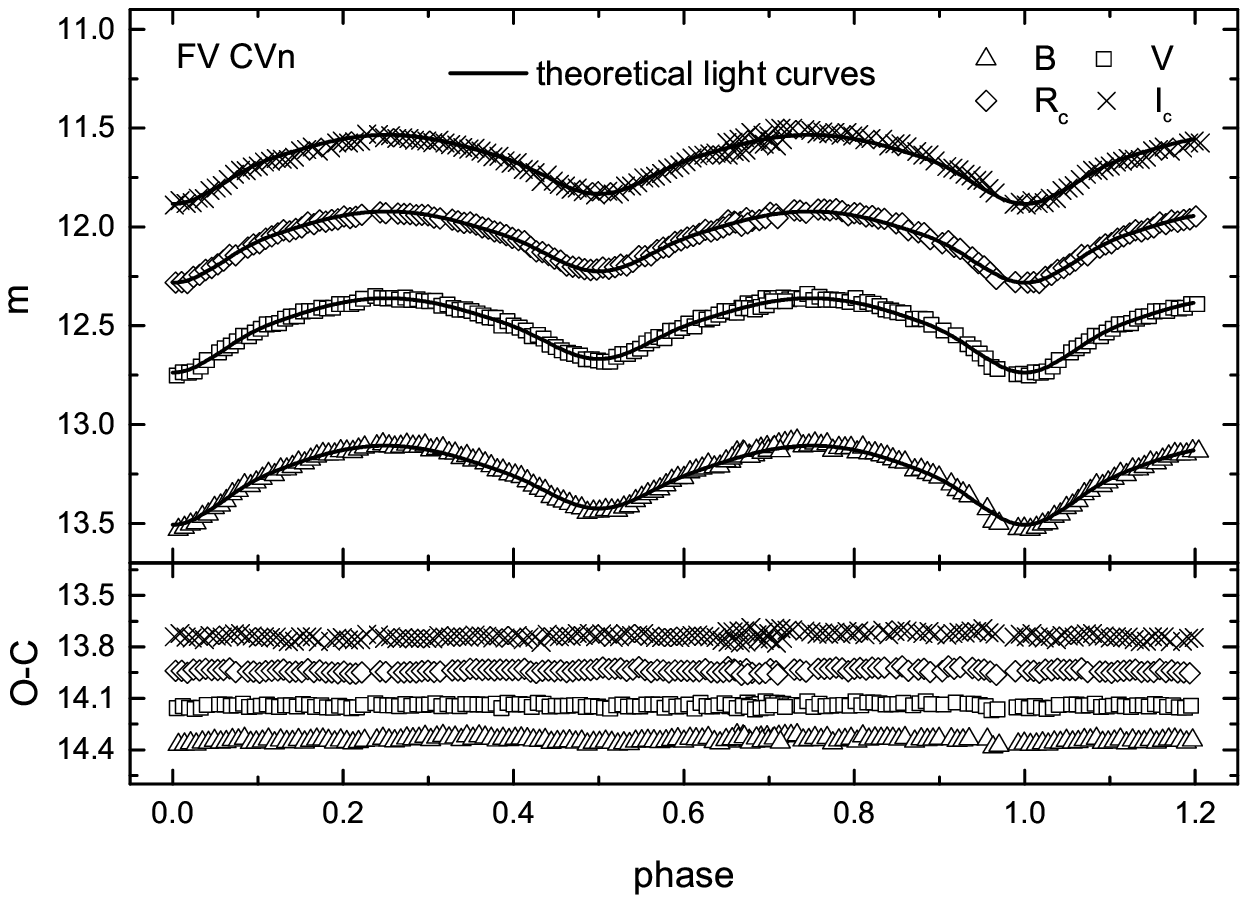}
\end{center}
\end{minipage}
\caption{The left panel is the $\Sigma$ - q curves of FV CVn. The small insert figure is an enlargement that q is from 0.9 to 1.5. The right panel is the agreement of FV CVn between observations and theoretical light curves derived by B, V, $R_c$, and $I_c$ light curves.}
\end{figure}

\begin{figure}
\begin{minipage}[c]{0.5\textwidth}
\begin{center}
\includegraphics[angle=0,scale=0.65]{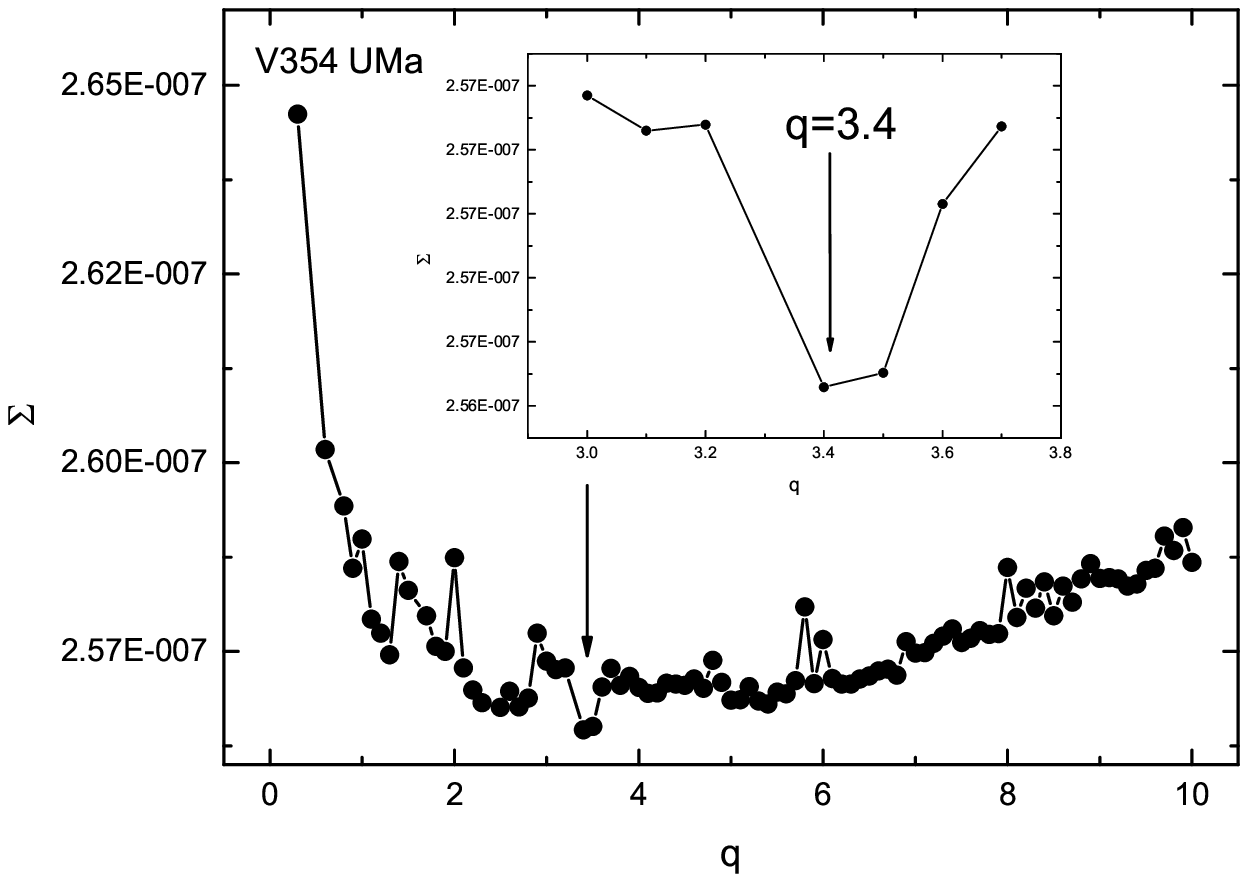}
\end{center}
\end{minipage}
\begin{minipage}[c]{0.5\textwidth}
\begin{center}
\includegraphics[angle=0,scale=0.63]{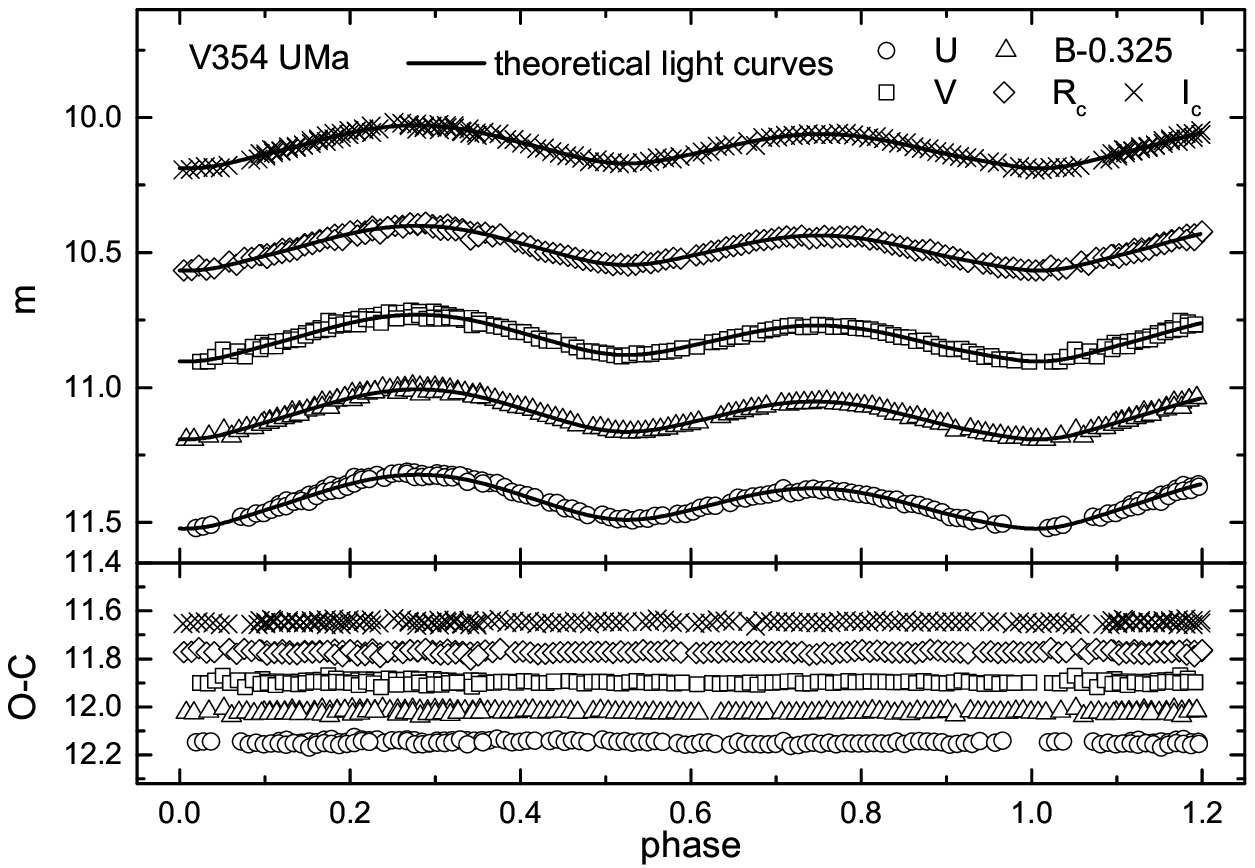}
\end{center}
\end{minipage}
\caption{The left panel is the $\Sigma$ - q curves of V354 UMa. The small insert figure is an enlargement that q is from 3.0 to 3.7. The right panel is the agreement of V354 UMa between observations and theoretical light curves derived by U, B, V, $R_c$, and $I_c$ light curves. It was computed with a dark spot on the secondary component.}
\end{figure}

\begin{table}
\scriptsize
\begin{center}
\caption{Photometric solutions of FV CVn, FP Lyn and V354 UMa}
\begin{tabular}{lccccccc}
\hline
Parameters           &\multicolumn{ 3}{c}{FP Lyn}  &\multicolumn{ 2}{c}{FV CVn}&\multicolumn{ 2}{c}{V354 UMa}\\
\hline
$T_1(K)$             &  5909     &  5909      &  5909      &   5470      &  5470       &  5841       &  5841        \\
                                                                               
$q(M_2/M_1) $        &  1.203(7) &  1.515(18) &  1.153(6)  &   1.071(10) &  1.032(25)  &  3.461(22)  &  3.622(40)   \\
$T_2(K) $            &  5688(15) &  5255(16)  &  5683(14)  &   5120(26)  &  5148(24)   &  5745(39)   &  5686(33)    \\
$i(deg)$             &  60.5(1)  &  59.3(1)   &  61.2(2)   &   64.2(2)   &  64.2(6)    &  50.4(7)    &  50.1(3)     \\
                                                                               
$ \Omega_{in} $      &  4.073    &  4.549     &  4.001     &   3.875     &  3.775      &  7.219      &  7.429       \\
$ \Omega_{out}$      &  3.513    &  3.970     &  3.445     &   3.325     &  3.230      &  6.594      &  6.802       \\
$ \Omega_1=\Omega_2 $  &  4.010(11)&  4.465(25) &  3.927(10) &   3.850(13) &  3.764(42)  &  7.174(41)  &  7.362(42)   \\
                                                                               
$L_{1U}/L_U$         &   -       &  -         &  -         &      -      &  -          &  0.271(9)   &  0.286(8)    \\
$L_{1B}/L_{B}$       &  0.517(2) & 0.599(3)   &  0.502(4)  &   0.597(4)  &  0.608(14)  &  0.264(6)   &  0.272(5)    \\
$L_{1V}/L_V$         &  0.502(2) & 0.554(3)   &  0.493(3)  &   0.572(4)  &  0.580(12)  &  0.258(5)   &  0.263(4)    \\
$L_{1R_c}/L_{R_c}$   &  0.495(2) & 0.530(2)   &  0.481(3)  &   0.557(3)  &  0.560(12)  &  0.256(4)   &  0.259(4)    \\
$L_{1I_c}/L_{I_c}$   &  -        & -          &  -         &   0.546(3)  &  0.551(11)  &  0.254(3)   &  0.256(3)    \\
$L_{3U}/L_{U}$       &  -        & -          &  -         &   -         &  -          &  -          &  -           \\
$L_{3B}/L_{B}$       &  -        & -          &  0.043(2)  &   -         &  0.000(4)   &  -          &  -           \\
$L_{3V}/L_{V}$       &  -        & -          &  0.032(1)  &   -         &  0.001(3)   &  -          &  -           \\
$L_{3R_c}/L_{R_c}$   &  -        & -          &  0.043(1)  &   -         &  0.004(3)   &  -          &  -           \\
$L_{3I_c}/L_{I_c}$   &  -        & -          &  -         &   -         &  0.005(3)   &  -          &  -           \\
                                                                               
$r_1(pole)$          &  0.348(6) &  0.331(1)  &  0.352(1)  &   0.354(1)  &  0.358(2)   &  0.262(1)   &  0.259(0)    \\
$r_1(side)$          &  0.366(6) &  0.347(1)  &  0.371(1)  &   0.373(1)  &  0.377(3)   &  0.273(2)   &  0.271(1)    \\
$r_1(back)$          &  0.402(8) &  0.384(1)  &  0.407(1)  &   0.406(1)  &  0.410(3)   &  0.308(2)   &  0.307(1)    \\
$r_2(pole)$          &  0.379(2) &  0.398(4)  &  0.375(2)  &   0.364(3)  &  0.362(8)   &  0.462(3)   &  0.468(3)    \\
$r_2(side)$          &  0.400(2) &  0.422(5)  &  0.396(2)  &   0.382(3)  &  0.381(10)  &  0.497(4)   &  0.506(4)    \\
$r_2(back)$          &  0.434(4) &  0.455(7)  &  0.430(3)  &   0.414(5)  &  0.413(15)  &  0.523(5)   &  0.533(6)    \\
$f$                  &  11(2)\%  &  14(4)\%   &  13(2)\%   &   5(3)\%&  2(8)\%     &  7(7)\%     &  11(7)\%     \\
\hline                                                                         
$spot$               & Star 1    &  Star 2    &  Star 1    &    -        &  -          &  Star 1     &  Star 2     \\
$colatitude(deg)$    & 52(2)     &  26(1)     &  52        &    -        &  -          &  47(4)      &  45(2)      \\
$longitude(deg)$     & 134(1)    &  98(4)     &  134       &    -        &  -          &  239(1)     &  244(2)     \\
$diameter(deg)$      & 22(0)     &  25(0)     &  22        &    -        &  -          &  17(0)      &  14(0)      \\
$T-factor$           & 1.14(00)  &  0.83(00)  &  1.14      &    -        &  -          &  1.29(1)    &  0.72(1)    \\
$\Sigma W(O-C)^2$    & $7.18\times10^{-8}$ & $9.44\times10^{-8}$ & $6.46\times10^{-8}$ &$7.26\times10^{-8} $ & $7.23\times10^{-8}$ & $1.20\times10^{-7}$ & $1.00\times10^{-7}$ \\
\hline
\end{tabular}
\end{center}
\end{table}

\begin{figure}\centering
\includegraphics[width=0.32\textwidth]{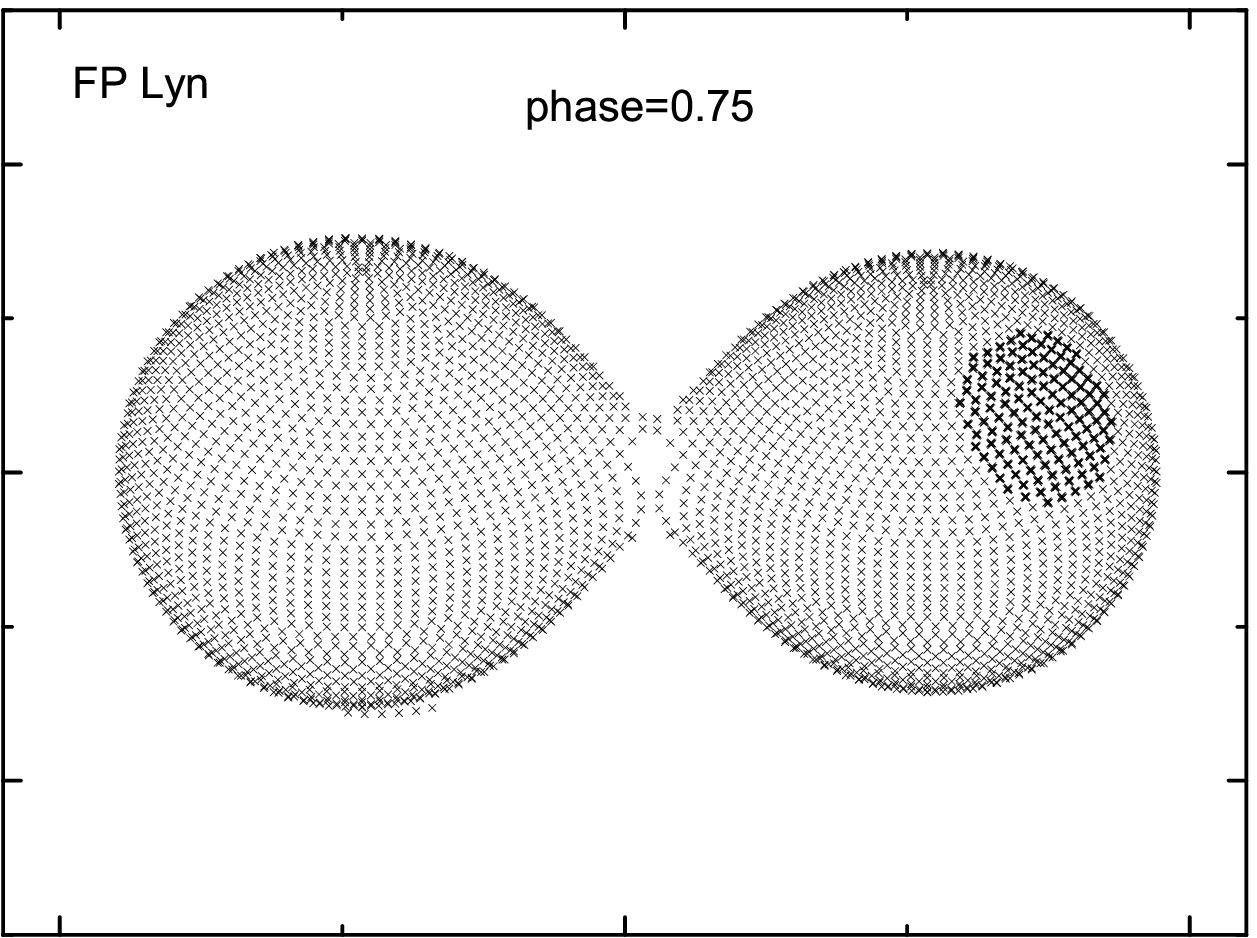}
\includegraphics[width=0.32\textwidth]{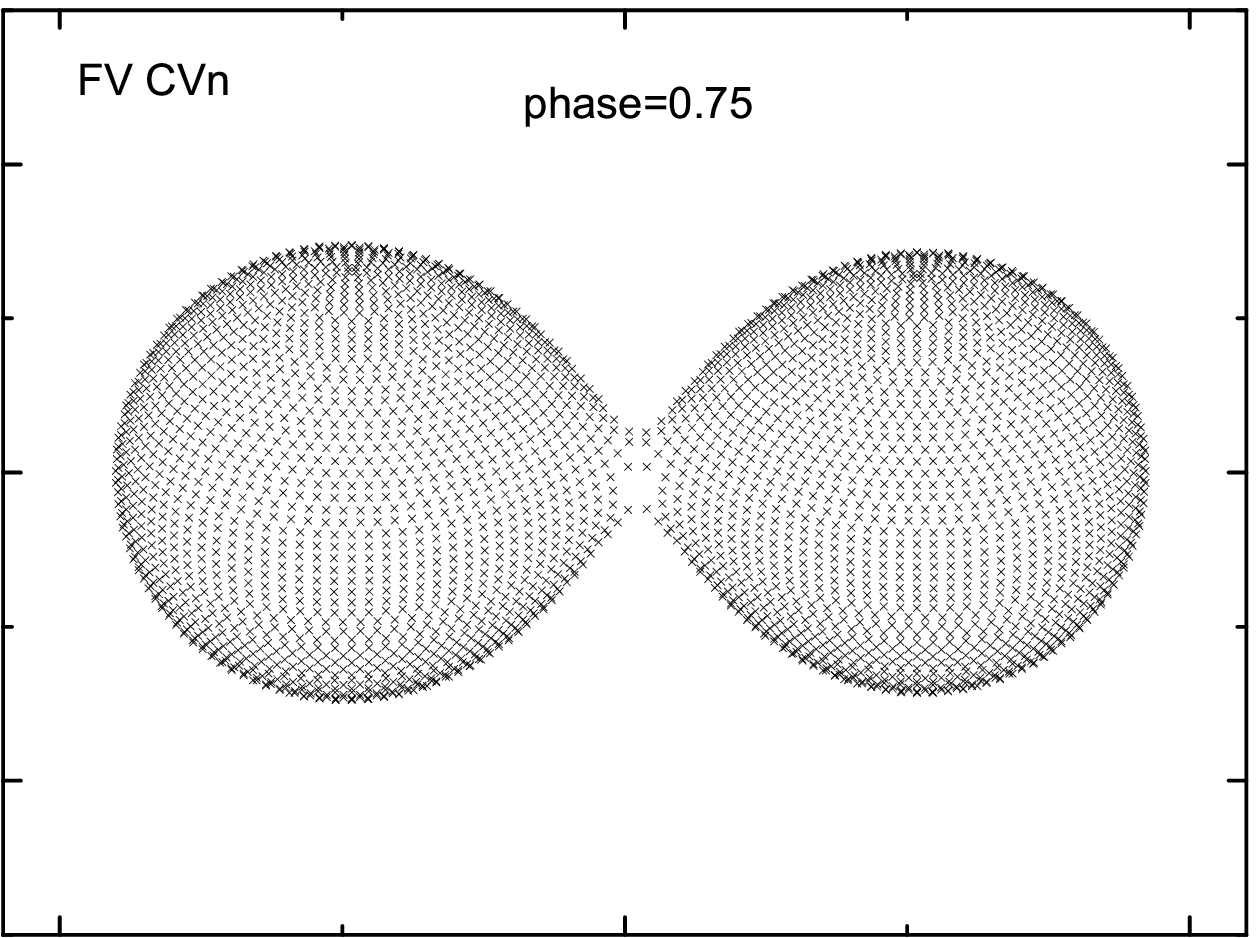}
\includegraphics[width=0.32\textwidth]{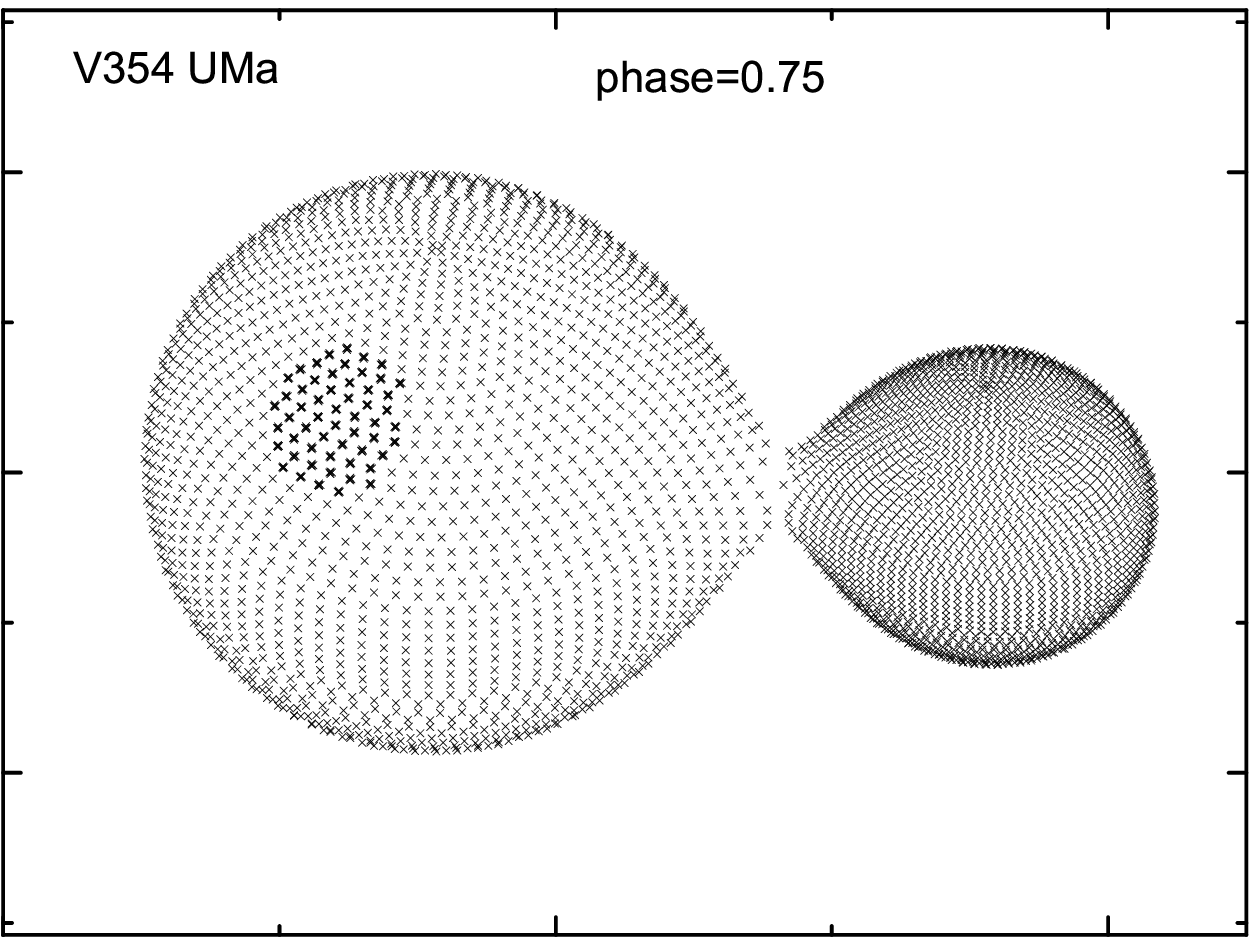}
\caption{Geometrical configuration of FP Lyn (left panel), FV CVn (middle panel) and V354 UMa (right panel) at phase 0.75.}
\end{figure}

\section{Study of orbital period}
There has been no analysis of orbital period since these three targets were discovered. Therefore, we analyzed all light minima what we could obtain. It has a total of 11 for FP Lyn, 34 for FV CVn and 13 for V354 UMa. All of these minima are listed in the first column of Tables 4, 5 and 6, respectively. The linear ephemeris to calculate the values of O-C are:
\begin{equation}
\textrm{FP Lyn: Min.I}=2453045.932+0.35909\textrm{E},
\end{equation}
\begin{equation}
\textrm{FV CVn: Min.I}=2453045.476+0.31537\textrm{E},
\end{equation}
\begin{equation}
\textrm{V354 UMa: Min.I}=2456008.526+0.293825\textrm{E},
\end{equation}
where the initial epoch and period are taken from O-C Gateway. Tables 4, 5 and 6 show the calculated O-C values of these three systems. The data corresponding to O-C values of FP Lyn and V354 UMa are plotted in Figure 6, and the data corresponding to $(O-C)_1$ values of FV CVn are plotted in Figure 7.

As Figure 6 shows, the overall trend of FP Lyn and V354 UMa show nonlinear variations. What we used to fit the O-C values are parabolas with upward opening. The least square method was used to revise ephemerides of FP Lyn and V354 UMa:
\begin{eqnarray}
\begin{split}
\textrm{FP Lyn: } Min.I = 2453045.9319850(\pm0.2580064) + 0.359082527\\
(\pm0.000000018)\times E + 2.061(\pm0.014) \times 10^{-10} \times E^{2},\\
\textrm{V354 UMa: } Min.I = 2456008.5273908(\pm0.4878153) + 0.2938251\\
(\pm0.000000048)\times E + 3.098(\pm0.065) \times 10^{-10} \times E^{2}.
\end{split}
\end{eqnarray}
In Equation (5), the coefficients of second-order term are positive, indicating that the period of FP Lyn and V354 UMa are all secular increase and the rates are calculated as $dp/dt=4.19(\pm0.03)\times10^{-7}$ d yr$^{-1}$ and $dp/dt=7.70(\pm0.16)\times10^{-7}$ d yr$^{-1}$, respectively. When we removed the parabolic values from O-C trend, the residuals are plotted in the down panel of Figure 6. We can see that the residuals are almost flat, indicating the O-C values of FP Lyn and V354 UMa were fitted well with the parabolic terms.

As Figure 7 shows, the data corresponding to the $(O-C)_1$ of FV CVn shows a downward parabolic component accompanied by a cyclic variation. Therefore, we used the following equation which is taken from Irwin (1952) to fit the overall trend,
\begin{eqnarray}
(O-C)_1=T_0+\Delta T_0+(P_0+\Delta P_0)E+{\beta\over{2}}E^2
+A[(1-e^2){\sin(\nu+\omega)\over(1+e\cos\nu)}+e\sin\omega] \nonumber\\
=T_0+\Delta T_0+(P_0+\Delta P_0)E+{\beta \over 2}E^2+A[\sqrt{(1-e^2 )}\sin E^*\cos\omega+\cos E^*\sin \omega],
\end{eqnarray}
where $\Delta T_0$ and $\Delta P_0$ are used to revise the initial ephemeris and the period, respectively. $\beta$ is the value of long-term change in period, $A$ and $e$ are the amplitude and eccentricity of light effect respectively, and $\omega$ represents the longitude of periastron. Then, all of the parameters in this equation were determined as listed in Table 7. The value of $\beta$ is negative which indicates that the variation of period has a secular decrease component with a rate of $dp/dt=-1.13(\pm0.06)\times10^{-6}$ d yr$^{-1}$. And if the long-term decrease was removed, the data of $(O-C)_2$ are plotted in the middle panel of Figure 7 and the periodic oscillation is seen clearly. The final fitting result shows that the amplitude and period of this oscillation are 0.0069 days and 10.28 years, respectively. When the cyclic oscillation were also removed, the residuals which were displayed in the lowest panel of Figure 7. All residuals seem to be near zero shows that almost no change can be found. Therefore, the data of $(O-C)_1$ were fitted well by using the Equation (6).

\begin{table}
\scriptsize
\begin{center}
\caption{Observed times and O-C values of minimum light for FP Lyn}
\begin{tabular}{lccccrrc}
\hline\hline
JD(Hel.)&Errors  &Type    &Min.&Epoch   &(O-C)   &Residuals   &References          \\
2400000+&        &        &    &        &        &            &                    \\
\hline
53045.9320  &            & ccd or pe  &  p  &  0        &  0.0000   &  0.0000    & (1)  \\
57101.4357  &   0.0041   &   pe    &  p  &  11294    &  -0.0023  &  -0.0006   & (2)  \\
57101.6149  &   0.0020   &   pe    &  s  &  11294.5  &  -0.0026  &  -0.0010   & (2)  \\
57122.4435  &   0.0004   &   ccd   &  s  &  11352.5  &  -0.0010  &  0.0006    & (3)  \\
57329.6355  &   0.0007   &   ccd   &  s  &  11929.5  &  -0.0010  &  -0.0008   & (3)  \\
57465.3743  &   0.0005   &   ccd   &  s  &  12307.5  &  0.0037   &  0.0029    & (3)  \\
57790.1618  &            &   ccd   &  p  &  13212    &  -0.0012  &  -0.0045   & (4)  \\
57790.3491  &            &   ccd   &  s  &  13212.5  &  0.0065   &  0.0032    & (4)  \\
57418.6891  &   0.0037   &   ccd   &  s  &  12177.5  &  -0.0005  &  -0.0009   & (5)  \\
57418.8717  &   0.0014   &   ccd   &  p  &  12178    &  0.0026   &  0.0022    & (5)  \\
57438.7979  &   0.0016   &   ccd   &  s  &  12233.5  &  -0.0004  &  -0.0010   & (5)  \\
\hline
\end{tabular}
\end{center}
\scriptsize
(1) GCVS; (2) H\"{u}bscher 2016; (3) Jury\v{s}ek et al. 2017; (4) VSOLJ 64; (5) This paper.
\end{table}

\begin{table}
\scriptsize
\begin{center}
\caption{Observed times and O-C values of minimum light for FV CVn}
\begin{tabular}{lccccrrrc}
\hline\hline
JD(Hel.)  &Errors &Method&Min.&Epoch  &    (O-C)$_1$ & (O-C)$_2$ & Residuals & References   \\
2400000+  &       &    &    &       &            &        &        &                \\
\hline
51251.8239  &  0.0007   & ccd&s & -5687.5 & -0.0137  & 0.0031   & -0.0031  & (1)    \\
51259.8720  &  0.0005   & ccd&p & -5662   & -0.0074  & 0.0093   & 0.0030   & (1)    \\
53045.4760  &  0.0009   & ccd&p & 0       & 0.0000   & -0.0068  & -0.0018  & (2)    \\
53060.4579  &  0.0013   & ccd&s & 47.5    & 0.0021   & -0.0049  & 0.0002   & (2)    \\
53060.6181  &  0.0020   & ccd&p & 48      & 0.0046   & -0.0024  & 0.0027   & (2)    \\
53068.4984  &  0.0018   & ccd&p & 73      & 0.0008   & -0.0063  & -0.0012  & (2)    \\
53081.4294  &  0.0019   & ccd&p & 114     & 0.0018   & -0.0053  & -0.0002  & (2)    \\
53094.3569  &  0.0012   & ccd&p & 155     & -0.0007  & -0.0079  & -0.0027  & (2)    \\
53094.5212  &  0.0021   & ccd&s & 155.5   & 0.0059   & -0.0013  & 0.0039   & (2)    \\
53117.3811  &  0.0017   & ccd&p & 228     & 0.0019   & -0.0056  & -0.0002  & (2)    \\
53117.5393  &  0.0011   & ccd&s & 228.5   & 0.0024   & -0.0050  & 0.0003   & (2)    \\
53464.4430  &  0.0020   & ccd&s & 1328.5  & 0.0046   & -0.0054  & 0.0009   & (3)    \\
53809.4517  &  0.0010   & pe &s & 2422.5  & 0.0040   & -0.0081  & -0.0025  & (4)    \\
53936.3887  &  0.0008   & pe &p & 2825    & 0.0066   & -0.0061  & -0.0015  & (5)    \\
54170.3973  &  0.0002   & pe &p & 3567    & 0.0143   & 0.0008   & 0.0023   & (5)    \\
54544.7391  &  0.0002   & ccd&p & 4754    & 0.0179   & 0.0035   & -0.0009  & (6)    \\
54682.3999  &  0.0011   & ccd&s & 5190.5  & 0.0219   & 0.0074   & 0.0017   & (7)    \\
54913.7168  &  0.0004   & ccd&p & 5924    & 0.0185   & 0.0040   & -0.0023  & (8)    \\
55000.4444  &  0.0003   & ccd&p & 6199    & 0.0208   & 0.0063   & 0.0001   & (9)    \\
55317.3859  &  0.0001   & ccd&p & 7204    & 0.0204   & 0.0065   & 0.0014   & (10)   \\
55317.5440  &  0.0001   & ccd&s & 7204.5  & 0.0209   & 0.0069   & 0.0019   & (10)   \\
55629.9079  &  0.0003   & ccd&p & 8195    & 0.0157   & 0.0028   & -0.0003  & (11)   \\
56001.8793  &  0.0006   & ccd&s & 9374.5  & 0.0141   & 0.0030   & 0.0027   & (12)   \\
56010.8612  &  0.0004   & ccd&p & 9403    & 0.0081   & -0.0029  & -0.0032  & (13)   \\
56031.3619  &  0.0004   & pe &p & 9468    & 0.0101   & -0.0008  & -0.0010  & (14)   \\
56034.3584  &  0.0004   & pe &s & 9477.5  & 0.0106   & -0.0003  & -0.0004  & (14)   \\
56069.5211  &  0.0036   & pe &p & 9589    & 0.0101   & -0.0006  & -0.0004  & (14)   \\
56074.7247  &  0.0003   & ccd&s & 9605.5  & 0.0102   & -0.0005  & -0.0003  & (12)   \\
57003.6183  &  0.0005   & ccd&p & 12551   & -0.0038  & -0.0066  & -0.0007  & (15)   \\
57119.8307  &  0.0002   & ccd&s & 12919.5 & -0.0034  & -0.0049  & 0.0013   & (16)   \\
57433.9306  &  0.0003   & ccd&s & 13915.5 & -0.0071  & -0.0047  & 0.0014   & (17)   \\
57842.9550  &  0.0002   & ccd&s & 15212.5 & -0.0110  & -0.0030  & -0.0003  & (18)   \\
58229.7505  &  0.0003   & ccd&p & 16439   & -0.0107  & 0.0034   & -0.0002  & (18)   \\
58229.9080  &  0.0003   & ccd&s & 16439.5 & -0.0109  & 0.0032   & -0.0004  & (18)   \\
\hline
\end{tabular}
\end{center}
\scriptsize
(1) Bl\"{a}ttler et al. 2004; (2) Diethelm 2004; (3) Diethelm 2005; (4) Diethelm 2006; (5) Diethelm 2007; (6) Nelson 2009a; (7) Diethelm 2009; (8) Nelson 2010; (9) Diethelm 2009b; (10) Dem\.{i}rcan et al. 2011; (11) Nelson 2012; (12) Diethelm 2012; (13) Nelson 6050; (14) H\"{u}bscher et al. 2013; (15) Jury\v{s}ek 2017; (16) Nelson 2016; (17) Nelson 2017; (18) This paper.
\end{table}

\begin{table}
\scriptsize
\begin{center}
\caption{Observed times and O-C values of minimum light for V354 UMa}
\begin{tabular}{lccccrrc}
\hline\hline
JD(Hel.)&Errors  &Type    &Min.&Epoch   &(O-C)   &Residuals   &References          \\
2400000+&        &        &    &        &        &            &                    \\
\hline
56008.5255 &  0.0015  &  pe    & p  & 0       & 0.0000    & -0.0019   & (1)             \\
56016.3144 &  0.0012  &  ccd   & s  & 26.5    & 0.0025    & 0.0006    & (2)             \\
56016.4583 &  0.0008  &  ccd   & p  & 27      & -0.0005   & -0.0024   & (2)             \\
56019.4003 &  0.0024  &  pe    & p  & 37      & 0.0033    & 0.0014    & (1)             \\
56019.5476 &  0.0018  &  pe    & s  & 37.5    & 0.0037    & 0.0018    & (1)             \\
57022.9616 &  0.0005  &  ccd   & s  & 3452.5  & 0.0053    & -0.0006   & (3)             \\
57023.1105 &  0.0010  &  ccd   & p  & 3453    & 0.0072    & 0.0013    & (3)             \\
57097.4494 &  0.0049  &  pe    & p  & 3706    & 0.0084    & 0.0019    & (4)             \\
57100.3855 &  0.0026  &  pe    & p  & 3716    & 0.0063    & -0.0002   & (4)             \\
57100.5326 &  0.0031  &  pe    & s  & 3716.5  & 0.0065    & -0.0001   & (4)             \\
57843.7671 &  0.0007  &  ccd   & p  & 6246    & 0.0107    & -0.0039   & (5)      \\
58231.7783 &  0.0004  &  ccd   & s  & 7566.5  & 0.0259    & 0.0055    & (5)      \\
58231.9164 &  0.0006  &  ccd   & p  & 7567    & 0.0171    & -0.0033   & (5)      \\
\hline
\end{tabular}
\end{center}
\scriptsize
(1) H\"{u}bscher et al. 2013; (2) G\"{u}rsoytrak et al. 2013; (3) Nelson 2015; (4) H\"{u}bscher 2016; (5) This paper.
\end{table}

\begin{figure}
\begin{minipage}[c]{0.5\textwidth}
\begin{center}
\includegraphics[angle=0,scale=0.65]{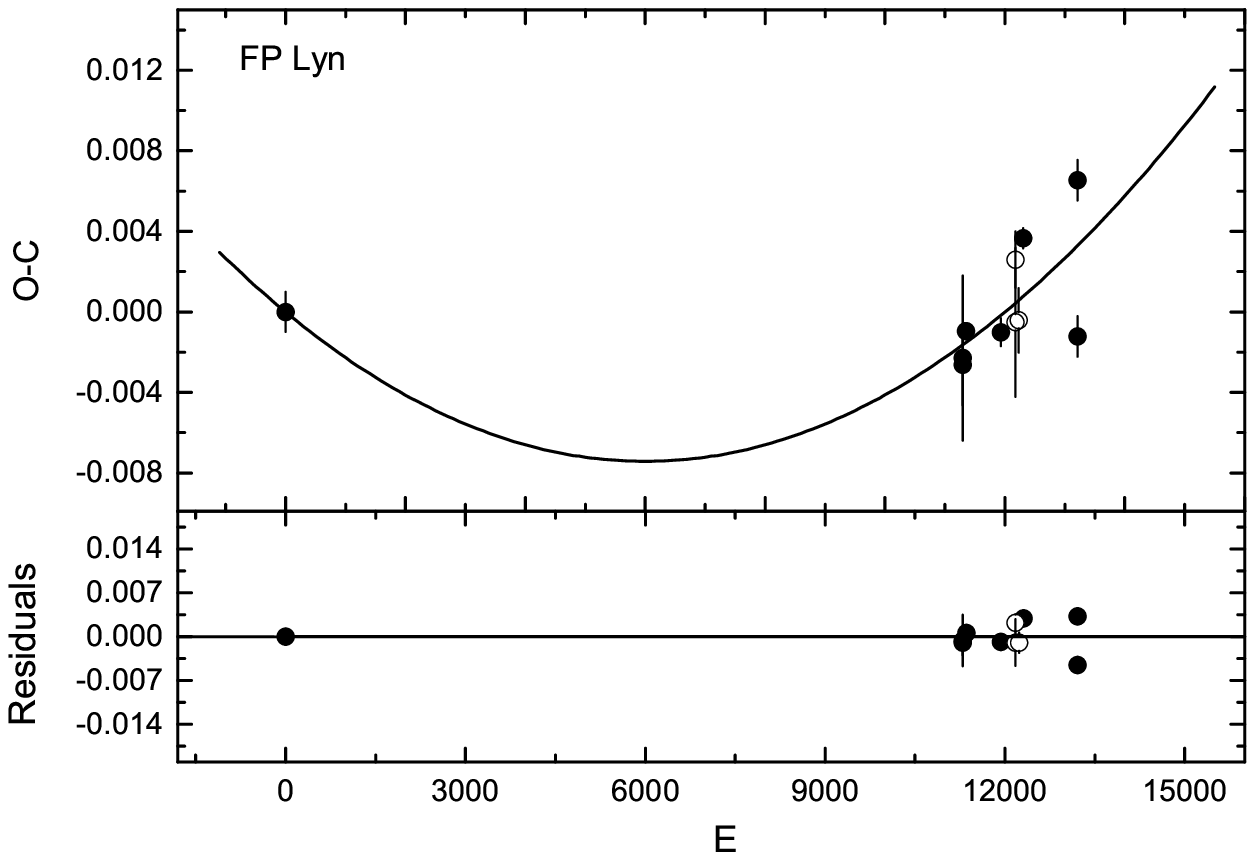}
\end{center}
\end{minipage}
\begin{minipage}[c]{0.5\textwidth}
\begin{center}
\includegraphics[angle=0,scale=0.63]{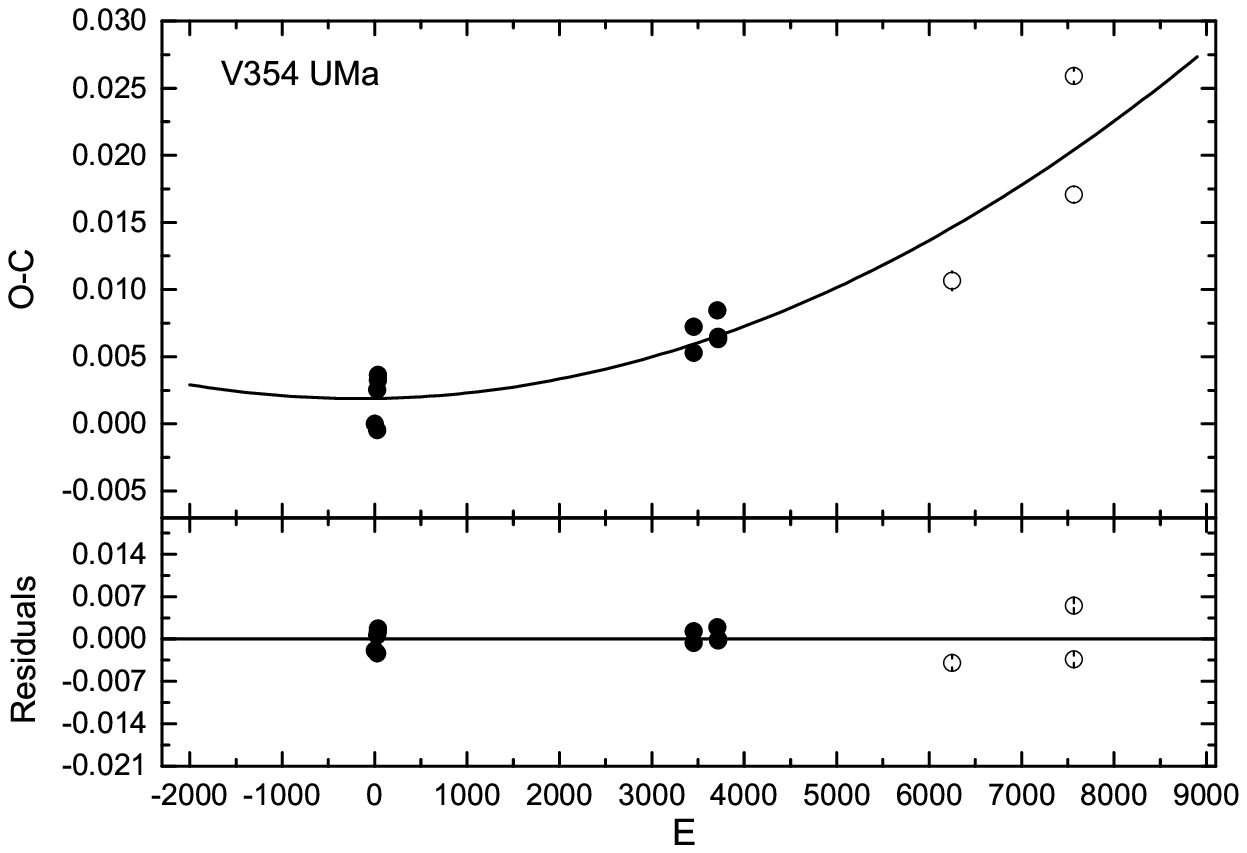}
\end{center}
\end{minipage}
\caption{$O-C$ diagram of FP Lyn (left panel) and V354 UMa (right panel). Top panel shows $O-C$ curve determined by the linear ephemeris of Equation (2) based on all available times of minimum light. The residual values which remove the quadratic term from the $O-C$ curve are plotted in the lower panel. Open circles refer to the light minima from our observations, while filled circles represent the data from other literatures.}
\end{figure}

\begin{figure}
\begin{center}
\includegraphics[angle=0,scale=1.2]{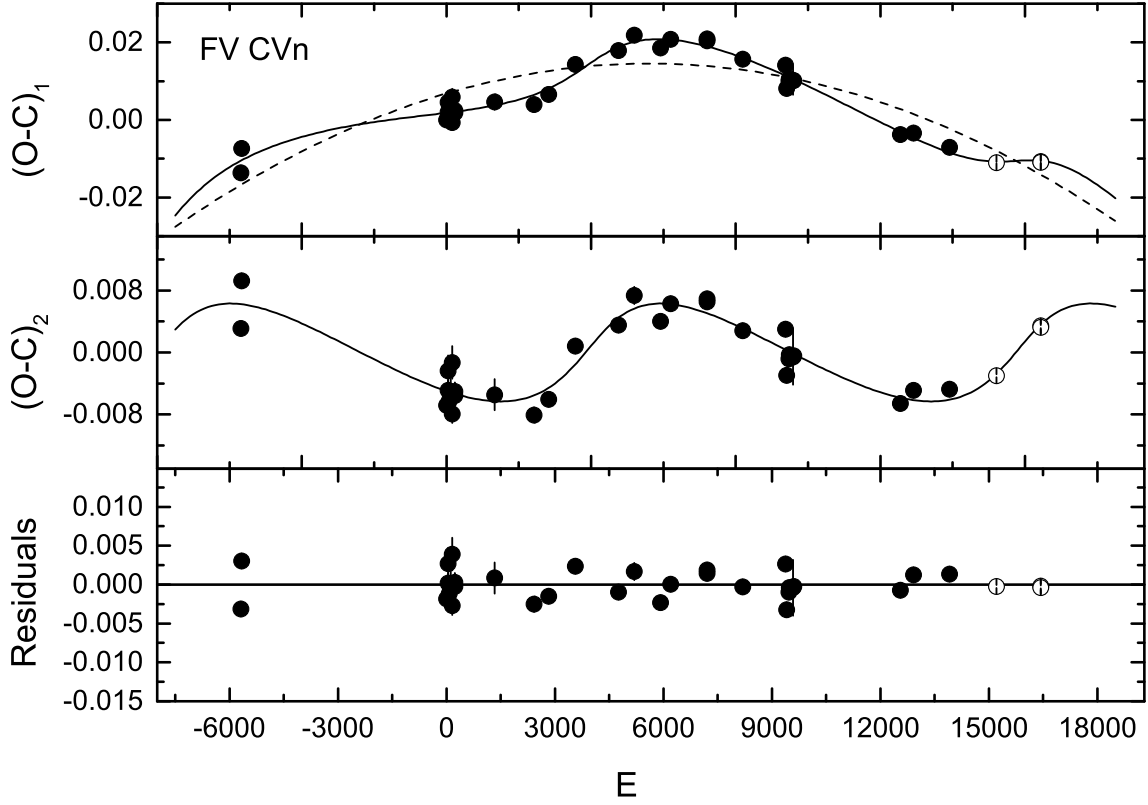}
\caption{$O-C$ diagram of FV CVn. Top panel shows $(O-C)_1$ curve determined by the linear ephemeris of Equation (3) based on all available times of minimum light. The $(O-C)_2$ values which remove the quadratic term from the $(O-C)_1$ curve are plotted in the middle. The residuals from full ephemeris of Equation (7) are displayed in the lower panel. Open circles refer to the light minima from our observations, while filled circles represent the data from other literatures. }
\end{center}
\end{figure}

\begin{table}
\begin{center}
\caption{Parameters for the fit of times of minimum light of FV CVn}
\begin{tabular}{lcc}
\hline\hline
Parameters &     Values &     Errors \\
\hline
$\Delta T_0$ (days)      &    0.00682            &  $\pm0.00064$  \\

$\Delta P_0$ (days)      & $2.75\times10^{-6}$ & $\pm0.16\times10^{-6}$  \\

$\beta$ (days $yr^{-1}$) & $-1.13\times10^{-6}$  & $\pm0.06\times10^{-6}$   \\

$A_{3}$ (d)              &     0.0069            &  $\pm0.0008$   \\

$e$                      &      0.41            &    $\pm0.16$   \\

$P_{3}$ (yr)             &     10.28             &   $\pm0.42 $   \\

$\omega (^\circ)$        &       9.2           &    $ \pm24.5 $   \\

$ T_P$ (HJD)             &  2422843.1            &   $\pm100244.6 $   \\
\hline
\end{tabular}
\end{center}
\end{table}

\section{Summary and conclusions}
Photometric solutions were determined by the light curves of multi-band by using the W-D program. Through the analysis of the final fitting parameters, we can obtain the following conclusions. All of these three systems FP Lyn, FV CVn and V354 UMa are shallow contact binaries with fill-out factors of $f = 13.4\pm1.9\%$, $f = 4.6\pm2.5\%$ and $f = 10.7\pm6.8\%$, respectively. Because the mass ratios of these three systems are all greater than 1, they are all W-subtype binary systems. And the mass ratios $q$ of FP Lyn, FV CVn and V354 UMa are $q = 1.153\pm0.006$, $q = 1.075\pm0.010$ and $q = 3.623\pm0.040$, respectively. As the spectra or the radial velocity curves of these three targets are all absent, the absolute parameters cannot be determined directly. The absolute parameters were determined by using the three-dimensional correlations of physical parameters for contact binaries taken from Gazeas (2009). The separation between two components were calculated using the Kepler's third law. Then, all of the absolute parameters of these three systems are listed in Table 8.

\begin{table}                  
\begin{center}                                                              
\caption{Absolute parameters of FP Lyn, FV CVn and V354 UMa}                            
\begin{tabular}{lcccccccc}                                                                          
\hline\hline                                                                                        
Objects	& \multicolumn{ 2}{c}{FP Lyn} & & \multicolumn{ 2}{c}{FV CVn} & &\multicolumn{ 2}{c}{V354 UMa} \\          
\cline{2-3}\cline{5-6}\cline{8-9}                                                         
Parameters & Values & Errors & & Values & Errors & & Values &Errors \\               
\hline                                                       
$M_1$ ($M_\odot$)  & 0.97     &    0.09   &  &  0.94   &    0.09   & &   0.29  &    0.05   \\         
$M_2$ ($M_\odot$)  & 1.11     &    0.11   &  &  1.01   &    0.10   & &   1.05  &    0.12   \\         
$R_1$ ($R_\odot$)  & 1.01     &    0.05   &  &  0.91   &    0.05   & &   0.60  &    0.04   \\         
$R_2$ ($R_\odot$)  & 1.07     &    0.05   &  &  0.94   &    0.04   & &   1.04  &    0.06   \\    
$L_1$ ($L_\odot$)  & 0.90     &    0.10   &  &  0.66   &    0.08   & &   0.36  &    0.05   \\         
$L_2$ ($L_\odot$)  & 1.02     &    0.12   &  &  0.71   &    0.09   & &   1.10  &    0.18   \\         
a($R_\odot$)       & 2.72     &    0.14   &  &  2.44   &    0.13   & &   2.05  &    0.14   \\         
\hline                                                                                               
\end{tabular}                                                                                        
\end{center}                                                                                          
\end{table}                                                                                         

\subsection{Mass Transfer and possible evolution}
From the analysis of orbital period of FP Lyn and V354 UMa, the overall trend of O-C show upward parabolic compositions with a rate of $dp/dt=4.19(\pm0.03)\times10^{-7}$ d yr$^{-1}$ and $dp/dt=7.70(\pm0.16)\times10^{-7}$ d yr$^{-1}$, respectively. The long-term increase can probably be explained by the mass transfer from the less massive star to the more massive star. Then, the following equation was used to calculate the rate of mass transfer of FP Lyn and V354 UMa,
\begin{eqnarray}
{\dot{P}\over P}=-3\dot{M_1}({1\over M_1}-{1\over M_2}).
\end{eqnarray}
Combining the parameters including mass, period and rate of period variation, the rate of mass transfer were determined as $dM_1/dt=-2.83\times10^{-6}\,M_\odot$ yr$^{-1}$ and $dM_1/dt=-3.50\times10^{-7}\,M_\odot$ yr$^{-1}$. The negative sign indicates that less massive component $M_1$ is losing mass, while the more massive component $M_2$ is receiving mass. As the mass ratio increases, so does the separation between two components. The degree of contact would decrease. They will evolve from the present contact binaries to the semi-contact or detached binaries. And this is not the end of evolution. According to the theory of stellar evolution, more massive component will fill its Roche lobe firstly. Therefore, the mass would transfer from the more massive star to the less massive one. Therefore, these two systems are good targets for studying the thermal relaxation oscillation (TRO) model (e.g., Lucy 1976; Flannery 1976; Robertson \& Eggleton 1977), and then we need to observe them continually.

The secular period decrease can be generally explained by the mass transfer and angular momentum loss (AML) by stellar wind. Firstly, assuming the long-term decrease of FV CVn is owing to the mass transfer from the more massive component to the less massive component. Then the rate of mass transfer was determined as $dM_1/dt=1.61\times10^{-5}\,M_\odot$ yr$^{-1}$ by using Equation (7). The positive sign indicates that the less massive component $M_1$ is receiving mass from $M_2$. Considering the mass of secondary star, the timescale of mass transfer was calculated as $\tau \sim {M_2\over \dot{M_2}}$ $\sim 6.28\times10^{4}$ yr, which is about $1\over764$ times of the thermal timescale of the secondary star (${GM^{2}\over RL} \sim 4.79\times10^{7}$ yr). Therefore, the mass transfer cannot be used to explain long-term change composition of FV CVn. The evolution of FV CVn is mainly driven by AML via magnetic braking. Then we collected other objects which are also shallow contact W-subtype eclipsing binaries, and their period show secular decrease. As shown in Table 9, the rate of long-term decrease of FV CVn is the most biggest among the objects, so we cannot exclude the possibility that it is a part of a long-term periodic change.

\begin{table}
\scriptsize
\begin{center}
\caption{W-subtype Shallow Contact Binaries with Decreasing Orbital Period}
\begin{tabular}{lccccccc}
\hline\hline
Star     &  Period(days)&  $q$   & $f(\%)$ &  $T_1(K)$ &  $T_2(K)$& $dp/dt$ (days yr$^{-1}$)&  References                          \\
\hline
MR Com   &  0.412746    &  3.912 & 10.0$\%$&  6530     &  6442    & $-5.3\times10^{-7}$     &  Qian et al. (2013b)                  \\
V1007 Cas&  0.332008    &  3.363 & 8.1$\%$ &  5135     &  4193    & $-1.8\times10^{-7}$     &  Li et al. (2018)                    \\
SS Ari   &  0.405991    &  3.250 & 9.4$\%$ &  5950     &  5745    & $-4.0\times10^{-7}$     &  Liu et al. (2009)                   \\
BX Peg   &  0.280422    &  2.660 & 14.6$\%$&  5887     &  5300    & $-2.1\times10^{-7}$     &  Li et al. (2015b), Lee et al. (2009) \\
V396 Mon &  0.396344    &  2.554 & 18.9$\%$&  6210     &  6121    & $-8.6\times10^{-8}$     &  Liu et al. (2011a)                  \\
LO Com   &  0.286361    &  2.478 & 3.2$\%$ &  5178     &  4875    & $-1.2\times10^{-7}$     &  Zhang et al. (2016)                 \\
VW Boo   &  0.342318    &  2.336 & 10.8$\%$&  5560     &  5198    & $-1.5\times10^{-7}$     &  Liu et al. (2011b)                  \\
GU Ori   &  0.470682    &  2.320 & 17.7$\%$&  6050     &  6021    & $-6.2\times10^{-8}$     &  Zhou et al. (2018)                   \\
TY Boo   &  0.317150    &  2.150 & 16.4$\%$&  5712     &  5108    & $-3.7\times10^{-8}$     &  Chiristopoulou et al. (2012)        \\
BM UMa   &  0.271220    &  1.852 & 17.0$\%$&  4982     &  4600    & $-9.8\times10^{-8}$     &  Samec et al. (1995)                 \\
V1139 Cas&  0.297101    &  1.583 & 3.6$\%$ &  6250     &  5933    & $-3.7\times10^{-7}$     &  Li et al. (2015c)                    \\
IK Boo   &  0.303117    &  1.146 & 2.2$\%$ &  5781     &  5422    & $-2.2\times10^{-7}$     & Kriwattanawong et al. (2017)\\
FV CVn   &  0.315365    &  1.075 & 4.6$\%$ &  5470     &  5120    & $-1.1\times10^{-6}$     &  This paper\\
\hline
\end{tabular}
\end{center}
\end{table}

\subsection{Are FP Lyn, FV CVn and V354 UMa all triple systems?}
As shown in Figure 7, after the downward quadratic term in the $(O-C)_1$ curves is removed, the $(O-C)_2$ data show it is a periodic oscillation. The amplitude and period of this sinusoidal variation are 0.0069 days and 10.28 yr respectively. The cyclic variation are generally caused by Applegate mechanism (Applegate 1992) or the light travel time effect (LTTE) by the third body.

Applegate mechanism (Applegate 1992) is that the magnetic activity makes the quadruple moment changing, then giving rise to variation of the orbital period. Assuming the magnetic cycle in the quadruple momentum of the primary component brings out the periodic oscillation. The following equation which was taken from Lanza \& Rodon\`{o} (2002) was used to calculate the quadruple momentum,
\begin{eqnarray}
{\Delta{P}\over P}=-9{\Delta Q\over Ma^2}.
\end{eqnarray}
The required variation of the quadruple moment for the primary star $\Delta Q=9.1\times10^{49}$ g cm$^2$ was determined using this equation, which is about two orders of magnitude smaller than the typical value $10^{51}$ to $10^{52}$ g cm$^2$. Therefore, Applegate mechanism cannot be used to explain the periodic variation.

The most possible reason leading to the cyclic variations in the $(O-C)_2$ curves of FV CVn is the light-travel time effect via the existence of a third body (e.g., Liao \& Qian 2010). There are many other examples to prove this view, such as BI Vul (Qian et al. 2013), TY UMa (Li et al. 2015a), V776 Cas (Zhou et al. 2016c) and UZ Leonis (Lee et al. 2018). Therefore, the following equation was used to calculate the parameters of the third body,
\begin{equation}
f(m)={(m_{3}\sin i^{'})^3\over (m_1+m_2+m_{3})^2}={4\pi\over GP^2_{3}}\times(a_{12}\sin i^{'})^3.
\end{equation}
$f_3(m)=1.62(\pm0.56)\times10^{-2}\,M_\odot$ of the additional component was determined. It is assumed that the orbital inclination of the third body $i^{'}$ and the orbital plane of the eclipsing binary FV CVn are coplanar ( $i^{'}= i =64^\circ.2$). Then the mass of the third body was obtained as $m_3=0.411(\pm0.086)\,M_\odot$ with a distance of $5.05(\pm1.20)$ AU. Therefore, assuming the third body is a main sequence star, then the luminosity was calculated as $L_3=0.04L_\odot$ according to Cox (2000), which accounts for about $2.5\%$ of the triple system of FV CVn. According to the previous photometric solutions, the ratio of the third light is too smaller,  it is just about 0.5\%. Although this also proves that the third light may exist, we still choose the result without the third light in the final fitting result. But FV CVn could still be a triple system.

Through analysis of the evolution and the presence of the third body of FV CVn, a comprehensive analysis of its evolution process was obtained. According to the statistical analysis of Stepie\'{n} \& Gazeas (2012), the low mass contact binaries (LMCBs) whose total mass is lower than 1.4 $M_\odot$ would be in pre-contact phase for a long time of 8-9 Gyr, nevertheless, the contact phase lasts for about only 0.8 Gyr. In addition, such systems show very short period (P $<$ 0.3 days) along with moderate mass ratios (0.2 $\leq q \leq$ 0.8). Thus the progenitor of FV CVn may be a short-period detached system with two main sequence stars, and the case of FV CVn is highly similar to that of BI Vul (Qian et al. 2013a). $f=4.6\%$ of fill-out factor supports further the assumption that it is a newly formed shallow contact binary and is in the initial state of contact state. From the orbital period analysis, no mass transfer occurred between two components on account of the timescale of mass transfer is greatly small. Therefore, it evolved into the presently shallow contact configuration due to the AML via magnetic braking. Admittedly, the third body played an important role for the evolution of FV CVn. The angular momentum may loss from the central pair to the additional component according to Kozai (Kozai 1962), as well as AML will cause the orbital period to decrease and lead to the contact degree increasing. Eventually, it will evolve into a deep contact phase (e.g., M4 V53, Li et al. 2017; NO Cam, Zhou et al. 2017) or merge into a rapidly rotating single star (e.g., Stepie\'{n} \& Gazeas 2012; Qian et al. 2013a). 

Although the orbital period analysis of FP Lyn and V354 UMa did not find any component of periodic change, the third light component was found in the photometric analysis of FP Lyn, and the proportion can not be ignored. This suggests that FP Lyn could be a triple system. Meanwhile we cannot rule out the possibility that V354 UMa is a triple system, maybe the third body of V354 UMa is too dim.

In summary, FP Lyn, FV CVn and V354 UMa are all W-subtype W UMa stars. The light curves of FP Lyn and V354 UMa are asymmetric, therefore, spots were added in the process of calculating parameters. The overall trend of O-C of FP Lyn and V354 UMa are all upward parabolic curves, revealing the period of both of them are prolonged increase. It may be explained by the mass transfer from the less massive star to the more massive star. While the $(O-C)_1$ curve of FV CVn shows an downward variation, after verification the long-term decrease cannot be explained by mass transfer, the AML due to magnetic activity may cause this phenomenon. The periodic oscillation of FV CVn is most likely to be caused by a third body through the analysis of orbital period and the photometric solutions with the added third light. In addition, the third light was also added in the photometric solutions of FP Lyn and V354 UMa, and finally we got a convergent solution for FP Lyn, while the convergent solution of V354 UMa was not found. This suggests that FP Lyn could be a triple system, therefore more light minima of these three systems are needed to obtain the O-C curves to analyze the period variation. Although the results we got are greatly reliable, the radial velocity curves can further confirm the obtained mass ratios and absolute parameters. Therefore, more observations combining photometry and spectra are needed for these three systems.

\section*{Acknowledgments}
RM acknowledges the financial support from the Universidad Nacional Aut\'onoma de M\'exico (UNAM) and DGAPA under grant PAPIIT IN 100918. This work is also supported by the Chinese Natural Science Foundation (No. 11703016), and by the Joint Research Fund in Astronomy (No. U1431105) under cooperative agreement between the National Natural Science Foundation of China (NSFC) and Chinese Academy of Sciences (CAS), by the Natural Science Foundation of Shandong Province (No ZR2014AQ019), and by Young Scholars Program of Shandong University, Weihai (No 20820171006), and by the Open Research Program of Key Laboratory for the Structure and Evolution of Celestial Objects (No. OP201704). The spectral data were provided by Guoshoujing Telescope (the Large Sky Area Multi-Object Fiber Spectroscopic Telescope LAMOST), which is a National Major Scientific Project built by the Chinese Academy of Sciences. Funding for the project has been provided by the National Development and Reform Commission. LAMOST is operated and managed by the National Astronomical Observatories, Chinese Academy of Sciences.

\end{document}